\documentclass[%
 preprint,
 amsmath,amssymb,
 aps,
]{revtex4-2}

\usepackage{graphicx}
\usepackage{dcolumn}
\usepackage{bm}
\usepackage{color}

\begin{document}

\title{Rotational symmetry breaking potential for two-dimensional magnets}

\author{Cl\'{a}udio J. DaSilva}
\email{claudio.silva@ifg.edu.br}
\affiliation{Instituto Federal de Goi\'as, Rua 76, Centro, Goi\^ania - GO, Brazil}
\author{L. S. Ferreira}
\affiliation{Instituto de F\'{i}sica, Universidade Federal de Goi\'{a}s, Av. Esperan\c{c}a s/n, 74.690-900, Goi\^{a}nia, GO, Brazil}%
\author{A. A. Caparica}
\affiliation{Instituto de F\'{i}sica, Universidade Federal de Goi\'{a}s, Av. Esperan\c{c}a s/n, 74.690-900, Goi\^{a}nia, GO, Brazil}%

\date{\today}

\begin{abstract}
Here we present a new perspective to the breakdown of ferromagnetic order in two-dimensional spin-lattice models employing the rotation of the underlying lattice. Using an Ising spin system on a square lattice as a prototype, we demonstrate that an additional low-symmetry interaction may lead to the absence of the truly long-range order and forms aperiodic structure, such as magnetic stripes. Employing annealing and entropic Monte Carlo simulations, we show that our model allows tuning between different phases, magnetically ordered as well as more exotic nonmagnetic phases such as Ising-nematic by changing only one control parameter, which is responsible for the arising of magnetic frustration. In addition, our methodology of considering the coupling between the magnetic structure and the host material can be extended to the study of any type of spin-exchange model in two dimensions and has many potential interesting ramifications and applications.
\end{abstract}

\maketitle

The last several years have witnessed a renewed interest in frustrated magnetism\cite{moessner2006,Lacroix2011,Kohamaetal2019,Kleinetal2019,Jinetal2012}, which was propelled by the discovery of actual compounds presenting new magnetic exotic phases which can be described by two-dimensional spin-exchange models\cite{Sapkotaelal2017}. These novel excitations inherent to frustrated systems include the possible quantum spin-liquid state\cite{Watanabe2016} and chiral ordered states\cite{Fert2017}. 

Experimentally observable, materials containing layers of transition metal ions on a square lattice are considered quasi-two-dimensional (quasi-2D) for magnetism, as the interactions between layers are much smaller than those within the layers. At sufficiently low energies their magnetic degrees of freedom can be mapped to effective spin-exchange models into 2D Bravais lattices. Commonly, the expected long-range magnetic order is suppressed by magnetic frustration, which arises when there is no unique ground state which simultaneously minimizes the energy of all exchange bonds. Generally, spin-exchange frustration can be due to either the geometry of the lattice (as in pyrochlore\cite{Sohnetal2017,Shinaokaetal2012}, spinel, or Kagome systems) or where the competing interactions between magnetic moments cannot be mutually satisfied (as the competing nearest-neighbor and next-nearest-neighbor interactions). In its turn, frustration enhances spin fluctuations, and novel complex phase diagrams, new emergent critical phenomena, and technical applications can emerge.

Nonetheless, mechanical vibrations of the crystal lattice (phonons) modulate the crystal electric field of the magnetic ion, thus inducing a direct relaxation between two different spin states, and unconventional magnetic properties could be linked to phonon renormalization. It has been known for a while that spin-phonon coupling induces complex collinear order in triangular antiferromagnets\cite{WangVishwanath2008} and noncollinear order in pyrochlore magnets \cite{Shinaokaetal2012,Sohnetal2017}. For strong spin-lattice coupling, striped collinear states can be observed in compounds such as $\alpha$NaFeO$_2$ and MnBr$_2$\cite{WangVishwanath2008}. Yet, despite the vast number of models describing spatial anisotropy of spin fluctuations\cite{Shannonetal2006,Jinetal2012,Guerreroetal2015}, a general lattice spin model considering the effects of spin-host lattice excitation coupling is still missing. 

Therefore, here we present a new perspective on the breakdown of ferromagnetic order in two-dimensional lattice spin models through the rotation of the underlying lattice. Our model adds a rotational symmetry-breaking potential native to the host material structure, which is arguably necessary to orient magnetic phases in macroscopic samples. We will show that such a model possesses ground states with various patterns such as stripes, checkerboards, and more complicated morphologies such as Ising-nematic. All these phases are tuned by changing only a single controlling parameter.

\begin{figure}[!t]
\includegraphics[scale=0.9]{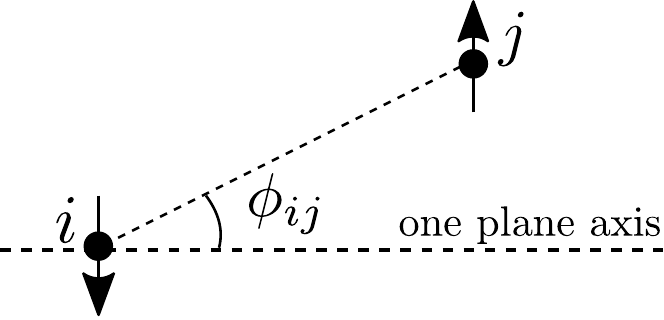}
\caption{\label{fig:scheme} A schematic illustration of the interaction symmetry.}
\end{figure}

Altogether, one can write a Hamiltonian for a model whose interaction depends on the symmetry of the layer as follows
\begin{align}
    \mathcal{H}=-\sum_{ij}J_{ij}^{\text{eff}}\vec{S}_i\cdot\vec{S}_j
    \label{hamiltonian}
\end{align}
with $J_{ij}^{\text{eff}}=J+J_a\cos{n\phi_{ij}}$ and $J$ is the regular magnetic exchange coupling. The constant $J_a$, determined by the strength of spin coupling to acoustic phonons, will be considered here as a phenomenological parameter. $\phi_{ij}$ is the angle between one of the plane axes and the line joining any two spins $i$ and $j$ as illustrated in Fig. \ref{fig:scheme}. $n$ is an integer number related to the degree of symmetry of the underlying crystal structure, namely $n=2$ for a tetragonal symmetry and $n=4$ for a square symmetry. Generally, the dependence of exchange on displacements may involve changes in the bond angles as well as
distances\cite{Bergmanetal2006,EdlundJacobi2010,WangVishwanath2008}. Such a theoretical procedure has already been successfully applied to strongly correlated system in Refs. \cite{sherman1995,claudio2010}. Besides, Ref. \cite{Sabiryanov1999} considered magnon-phonon interactions in Iron and observed that the effect of phonons on magnons depends strongly on the symmetry of the underling lattice. Also, Ref. \cite{Junqietal2013} proposed a Heisenberg model that couples the spin and lattice degrees of freedom for BCC iron. They have demonstrated by using Monte Carlo simulations that a higher transition temperature is obtained when a background contribution from phonons is considered. 

In the following, we will consider a single square lattice with only Ising spin states $\sigma=\pm 1$ with ferromagnetic and the spin-phonon interactions given by Eq. \ref{hamiltonian}. Hence, the dimensionless interaction energy will be $-\sum_{\langle ij \rangle}(1+\kappa\cos{2\phi_{ij}})\sigma_i\sigma_j$ for the tetragonal symmetry and $-\sum_{\langle ij \rangle}(1+\kappa\cos{4\phi_{ij}})\sigma_i\sigma_j$ for the square symmetry, where $\kappa=J_a/J$. $\langle ij \rangle$ indicates that the sum runs over all pairs of nearest-neighbor spins. Here, the parameter $\kappa$ quantifies the level of magnetic frustration. Although the geometry of the square lattice does not lead to frustration, the relative strengths of $J$ and $J_a$ determine the magnetic ground state, and a frustrated state can arise from their competition. The case of a square lattice deserves our initial attention since recent inelastic neutron scattering measurements confirmed an effective one-dimensional coupling in the highly frustrated itinerant magnet CaCo$_{2-y}$As$_2$\cite{Sapkotaelal2017} where the authors have identified Néel-(or checkerboard)type antiferromagnetic (AFM), stripe-type AFM, and ferromagnetic (FM)-type ordering occurring and been suppressed near the region of maximum frustration. Remarkably, within our approach we were also able to identify all those phases including a spin-nematic state at the maximum frustration which occurs when $\kappa=\pm 1$ for $n=2$ and $\kappa=-1$ for $n=4$.

First, we focus on the zero-temperature phase diagram as a function of the single parameter $\kappa$ obtained by two different numerical simulations techniques, namely Monte Carlo annealing and entropic sampling. Simulated annealing (SA) on each size was done using a uniform slow annealing schedule as follows: The simulation consists of $M$ segments of $N$ Monte Carlo steps each so that the total number of Monte Carlo steps is $NM$. The inverse temperature is increased after each Monte Carlo step as $\beta=1/k_BT=\beta_0+i\Delta\beta_M \quad (i=1,...,NM)$, where $\Delta\beta_M=(\beta_M-\beta_0)/NM$, with $\beta_0=0.0,\beta_M=8.0, M = 300$ and $N = 10^6$ steps of single spin flips updates with Metropolis dynamics. We executed ten independent annealing runs to ensure the stability of the results.

By its turn, the entropic Monte Carlo simulations, based on a modified Wang-Landau (WL) scheme\cite{WangLandau2001b}, are mainly applied to study the thermodynamics of the system at the critical values of $|\kappa|=1$. In the WL method, a random walk is performed in the energy space, which allows estimating the density of states $g(E)$, while a flat histogram for the energy distribution is generated. With the density of states in hands, one can calculate the canonical averages of any thermodynamic quantities. In our simulations, some easily implementable changes to the method are included, which lead to improved accuracy and to noticeable savings in CPU time. In particular,  ($i$) the density of states is updated only after every Monte Carlo sweep, such that we discard very correlated configurations, ($ii$) the microcanonical averages are accumulated beginning from the eighth WL level ($f_7$), so that we avoid considering the initial configurations that do not match with those of maximum entropy\cite{Caparica2012}, ($iii$) a checking parameter $\varepsilon$ is used for halting the simulation \cite{Caparica2014} (the integral of the specific heat over a range of temperatures  is calculated using the current density of states during the simulations and the simulations are halted if this quantity varies less then $10^{-4}$ during a whole WL level), and ($iv$) a single run is performed for all lattice sizes up to the Wang-Landau level $f_6$ and then, the further simulations begin from these outputs, since up to this point the current density of states is not biased yet and the results we reach are similar to those that would be obtained beginning from the first WL level $f_0$\cite{Ferreira2018}, a measure that saves about 60\% in computational time.
\begin{figure}[!t]
\includegraphics[scale=1.0]{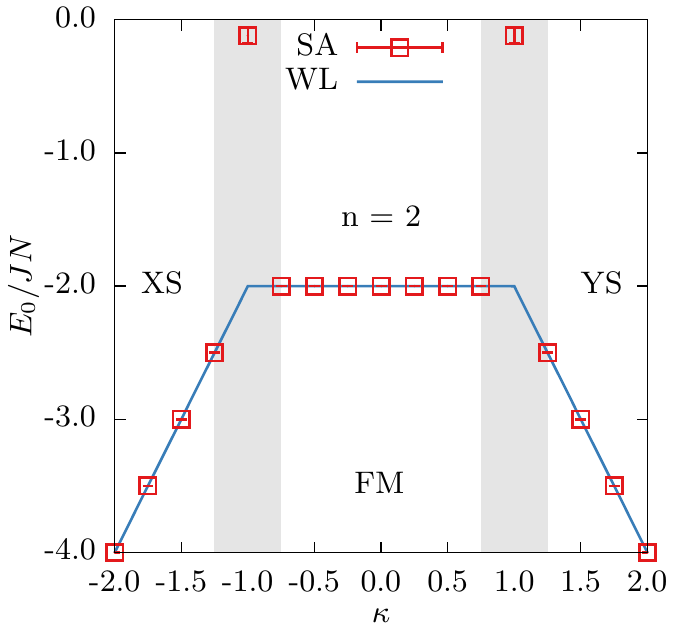}
\includegraphics[scale=1.0]{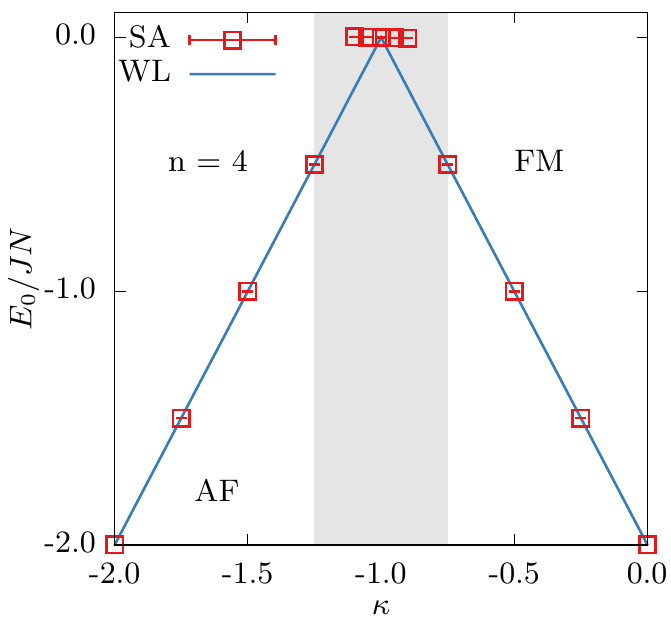}
\caption{Zero-temperature phase diagram for the symmetries $n=2$ and $4$ obtained using both WL and SA procedures. For $n=2$, on increasing the frustration parameter $\kappa$, an stripe XS state is followed first by a fully frustrated state (shaded area), second by a  ferromagnetic FM state, third by another fully frustrated region, then an stripe YS state. For $n=4$ an antiferromagnetic/checkerboard (AF) state is followed first by a fully frustrated state and then a ferromagnetic (FM) state.}
\label{ground}
\end{figure}

The phase diagrams shown in Fig. \ref{ground} were obtained using lattices with periodic boundary condition and size $42\times 42$, and choosing the state with the lowest energy per site $E_0/JN$. With our numerical procedures, we have identified four (two) distinct ordering phases of the ground states for $n=2$ ($n=4$). For $n=2$, the stripes morphology -- i.e., magnetic domains displayed in stripes of alternating spins -- becomes energetically favorable compared to the uniform ferromagnetic when $|\kappa|> 1$. As the value of the frustration parameter increases, the stripe XS phase is followed first by a typical Ising ferromagnetic FM state (with $E_0/JN=-2.0$), and then a stripe YS state. For $n=4$, a checkerboard pattern associated with an antiferromagnetic AF ground state is identified in the region where $\kappa < -1$ followed by a ferromagnetic FM state for $\kappa > -1$. More precisely, SA results reveal a frustration region (shaded area) when $0.75< |\kappa|<1.25$ with a maximum frustration at $|\kappa|=1$. In this region the stripe width $h$ changes with $\kappa$. In this range of $\kappa$, striped configurations of width $h>1$ coexist with the ferromagnetic state and a first-order phase transition seems to take place, as we shall see later. While the SA algorithm does not guarantee convergence to the ground state (as one can see the system get stuck in a value of $E_0/JN$ close to zero for $|\kappa|=1$.), the WL procedure always finds an ordered phase even in the frustrated region where highly degenerated ground states emerge. We nevertheless believe that an accurate picture emerges from the SA procedure. A frustration region is also identified for $n=4$ case at $-1.25<\kappa<-0.75$. However, the maximum frustration condition produces a system of non-interacting spins since $E_0/JN=0.0$. Hence, a thermodynamical analysis in this case will become complete when considering external fields coupled to all the spins, which is out of the scope of the present paper.
\begin{figure}[!t]
\includegraphics[scale=0.54]{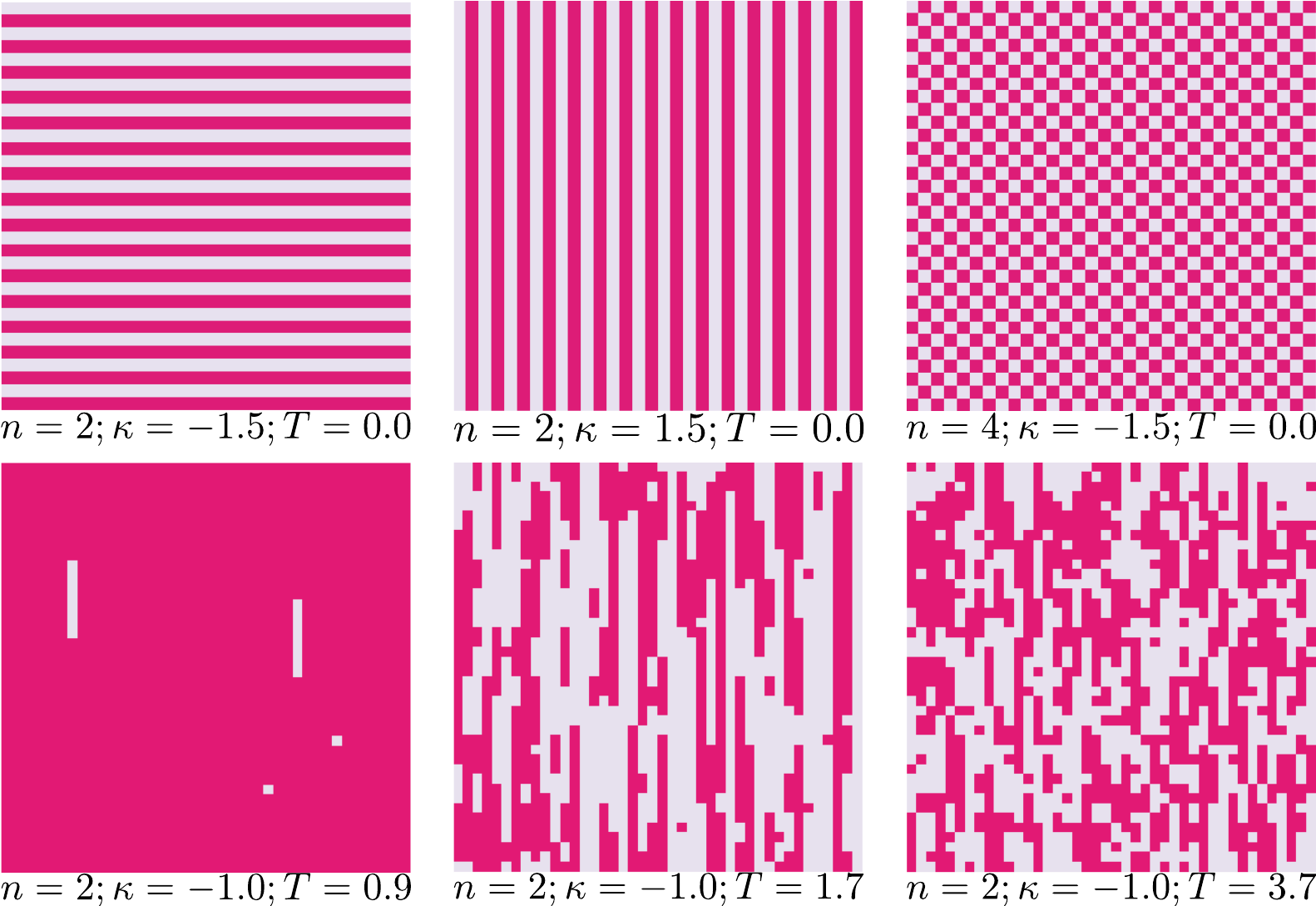}
\caption{Typical equilibrium spin configurations of the ground state (top) for both symmetries and close to an order-order transition (bottom left) for $n=2$. The presence of high density of topological defects, mainly dislocations in the directions of the underlying striped structure (bottom middle) is in agreement with the qualitative description of the Ising nematic phase.}
\label{confs}
\end{figure}

Fig. \ref{confs} shows typical equilibrium spin configurations of the non FM ground state patterns, namely stripes and checkerboard, for both symmetries obtained after a WL simulation\cite{Ferreira2021}. Also, it is shown some configurations in the region of maximum frustration for $n=2$. At the maximum frustration and low temperatures, stripes and ferromagnetic phases coexist. As the temperature increases, the striped structure acquires several dislocations in the directions of its underlying structure. The presence of such defects reduces the orientational order, which is in agreement with the qualitative description of the Ising nematic phase\cite{Cannas2006}. With a further increase of the temperature, domains with predominantly square corners pop out before the system reaches the paramagnetic state. Such a phase is known as a tetragonal liquid\cite{Cannas2004}.

To better characterize the different ordering phases and possibly the order-disorder phase transition proper order parameters were assigned to different values of $\kappa$. The XS and YS stripe phases observed for $n=2$ are characterized by a two-component order parameter $m_{stp}^2=m_i^2+m_j^2$ with $m_i=N^{-1}\sum_{i=1}^L\sum_{j=1}^L\sigma_{i,j}(-1)^{i}$, $m_j=N^{-1}\sum_{i=1}^L\sum_{j=1}^L\sigma_{i,j}(-1)^{j}$,
and $(i,j)$ is the coordinate of a site on the lattice\cite{Jinetal2012}. Hence, the stripe susceptibility is defined as $\chi_{stp}=N(\langle m_{stp}^2\rangle-\langle |m_{stp}| \rangle^2)/T$. For $n=4$ case and $\kappa<-1$, a staggered magnetization $m_{stg}=N^{-1}\sum_{i=1}^{L}\sum_{j=1}^L \sigma_{i,j}(-1)^{i+j}$ and consequently the staggered susceptibility $\chi_{stg}=N(\langle m_{stg}^2\rangle-\langle |m_{stg}| \rangle^2)/T$. For both symmetries in the region of maximum frustration, we defined an order parameter as following $m_\kappa = \sum_{j=1}^L\lvert\sum_{i=1}^L\sigma_{i,j}\rvert$ with the corresponding susceptibility $\chi_\kappa=N(\langle m_{\kappa}^2\rangle-\langle |m_{\kappa}| \rangle^2)/T$. This order parameter is able to capture emergent order induced by the second interaction term in Eq. \ref{hamiltonian} and is alone the proper way to reveal an order-disorder phase transition in a region where several ordered phases coexist.
\begin{figure}[!t]
\includegraphics[scale=0.8]{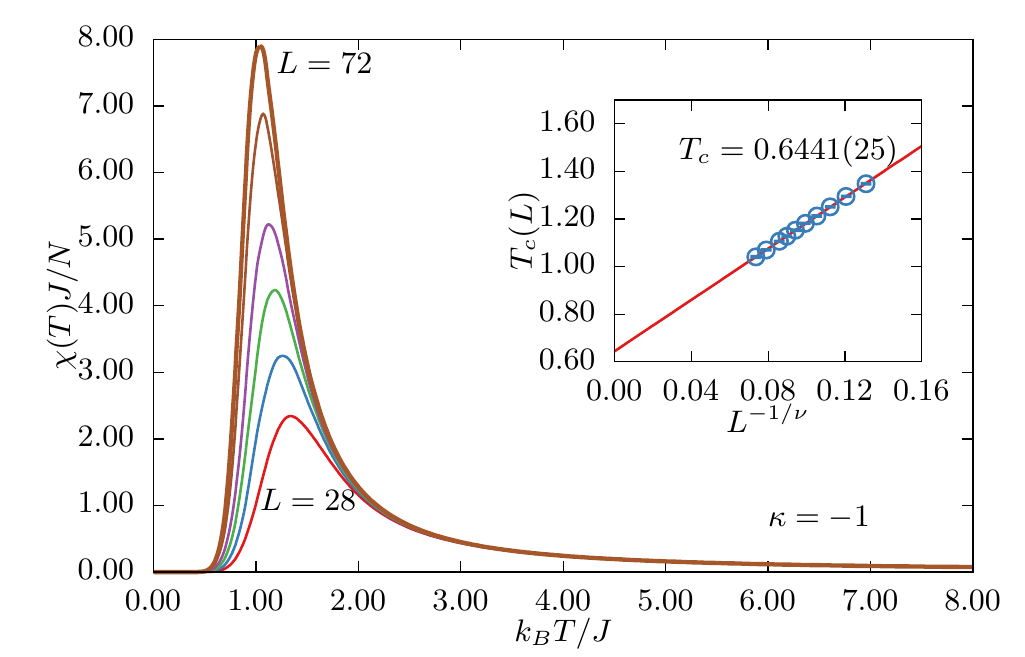}
\caption{Scaling behaviour of the susceptibility for $n=2$ symmetry at the maximum frustration ($\kappa=-1$). The inset shows a FSS analysis to obtain the transition temperature.\label{fss}}
\end{figure}

In this paper, we are not going to concentrate on the thermodynamics of the phase transitions in the range of the $\kappa$ out of the frustration region. We aim at the determination of the critical behavior only at the maximum frustration. For the other values of $\kappa$ we calculate the above order parameters only to obtain the full $T_c\times \kappa$ phase diagram for both symmetries. To perform the finite-size scaling (FSS) analysis at the maximum frustration, we used the outcome of the entropic simulations based on the WL algorithm. We ran sampling for lattice sizes of $L= 28$, 32, 36, 40, 44, 48, 52, 56, 64 e 72 with $N=24$, 24, 24, 20, 20, 20, 16, 16, 16, and 12 independent runs, respectively. The flatness criterion used was 80\%. Every run was stopped at the twentieth level of the WL, $f_{21}$, where the parameter $\varepsilon \sim 10^{-4}$. The order parameter $m_{\kappa}$ was certain in characterizing an order-disorder phase transition at the maximum frustration, as can be seen by the behavior of the susceptibility $\chi_\kappa$ shown in Fig. \ref{fss}. To obtain the critical temperature, we first estimate the value of the critical exponent $\nu=0.6101(16)$ using the quantities defined from the derivatives of the logarithm of the order parameter $m_\kappa$\cite{Caparicaetal2000}. Then, we used the scaling law $T_c(L) = T_c+aL^{-\frac{1}{\nu}}$, where $T_c$ is the critical temperature of the infinite system. In the inset of Fig. \ref{fss} we show the expected FSS behavior for a second-order phase transition and the linear regression yields $T_c=0.6441(25)$ for $n=2$. For $n=4$ at the maximum frustration, since the system behaves as an ensemble of non-interacting spins, the expected critical temperature is exactly $T_c=0.0$.
\begin{figure}[!t]
\includegraphics[scale=1.0]{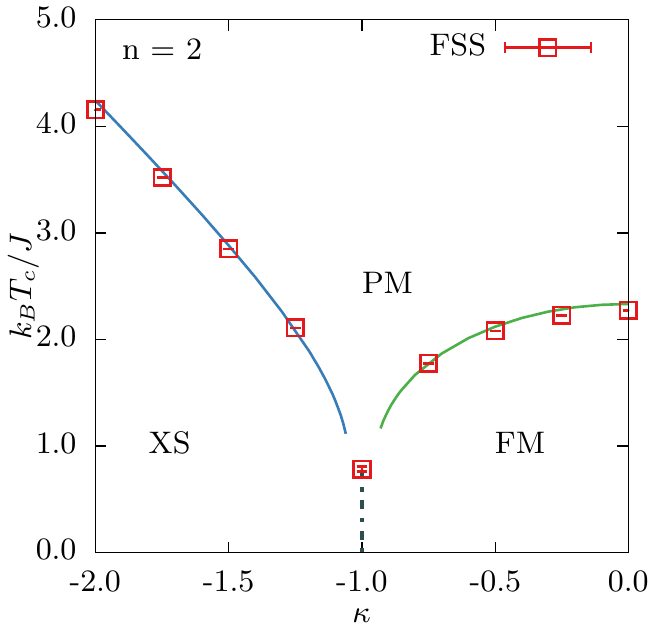}
\includegraphics[scale=1.0]{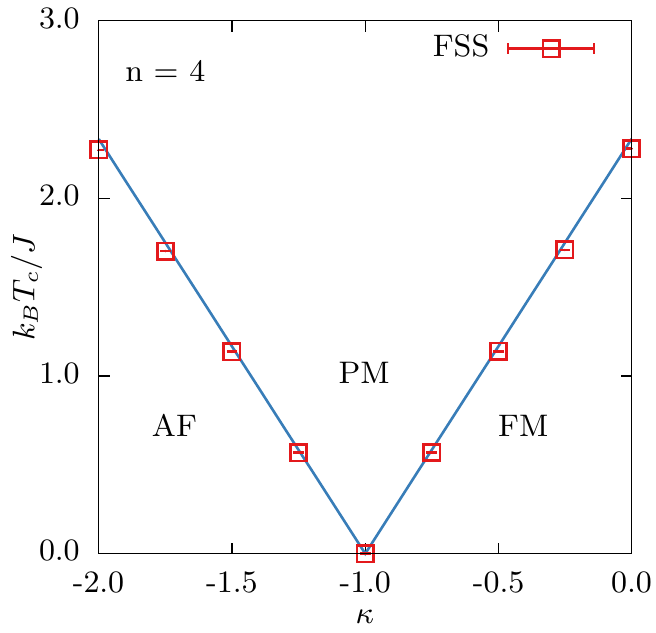}
\caption{Phase diagram in the coupling-temperature plane for both symmetries. Dashed and solid curves indicate first-order and continuous phase transitions, respectively. The square symbols indicate the critical points after a FSS analysis.\label{pdiagram}}
\end{figure}

To construct the $T_c\times \kappa$ phase diagrams we estimate the joint density of states $g(E_1,E_2)$, where $E_1$ and $E_2$ are, respectively $E_1=\sum_{\langle ij\rangle} \sigma_i\sigma_j$ and $E_2=\sum_{\langle ij\rangle} \sigma_i\sigma_j\cos n\phi_{ij}$\cite{Ferreira2019}. Hence, the thermodynamic averages can be obtained for any value of the parameters $T$, $J$, and $J_a$\cite{Ferreira2021,Jorge2021}. For these cases we used the lattice sizes $L=16$, 20, 24, 28, 32, 36 e 40. Due to the high computational cost, the WL runs were done up to $f=f_{15}$. Fig. \ref{pdiagram} shows phase diagrams for both $n=2$ (top) and $n=4$ (bottom). The solid line represents the temperature of the maximum of the susceptibility for $L=32$. The symbols with the corresponding error bars are the FSS analysis for different values of $\kappa$. For $n=2$, the discontinuity of the solid line indicates the limit of validity  of the order parameters $m_{stp}$ as $\kappa\rightarrow(-1)^-$ and $m$ as $\kappa\rightarrow(-1)^+$. The point at the maximum frustration $\kappa=-1$ was obtained by an FSS analysis of the $m_\kappa$ order parameter. In the region of phase coexistence (dashed line in Fig. \ref{pdiagram}), the energies of the ordered phases are too close and there is a huge number of configurations with even closer energies. Hence, small thermal fluctuations is enough to allow low energy configurations to be filled. That is the reason for critical temperatures in the frustrated region be close to zero. For $n=4$, as expected the lines of phase transition follow the evolution of the ground state energy $E_0(\kappa)$ as a function of $\kappa$ and the transition temperature at the maximum frustration occurs only at $T=0$. While the lines of phase transitions shown are related to an order-disorder transition, it is also expected an Ising-nematic phase between the stripes and disordered phases\cite{Guerreroetal2015,Cannas2004}. However, an analysis of the nature of such transition is left to be dealt with in future work.

In summary, we have proposed a simple spin lattice model that couples the spin degree of freedom to the underlying lattice symmetry. Our model allows tuning between different phases, magnetically ordered as well as more exotic nonmagnetic phases by changing only one control parameter. Also, it allows the manifestation of a spin-nematic state at the maximum frustration coexisting with two other ordered phases. In the present form, our model may offer a generic explanation as to why exotic spin patterns are so frequently observed in various natural magnetic systems. Besides, our results motivate further systematic studies including spin systems with 2D (XY) or 3D (Heisenberg) degrees of freedom, other lattice geometries such as honeycomb, triangular, or Kagome as well as systems with long-range interactions.

We thank the computer support from LaMCAD/UFG.

\bibliography{aising}

\end{document}